# VR Sickness Prediction from Integrated HMD's Sensors using Multimodal Deep Fusion Network

**Rifatul Islam**\*
Department of Computer Science
The University of Texas at San Antonio
San Antonio, TX 78249

**Kevin Desai**†
Department of Computer Science
The University of Texas at San Antonio
San Antonio, TX 78249

**John Quarles**‡
Department of Computer Science
The University of Texas at San Antonio
San Antonio, TX 78249

August 14, 2021

## Abstract

Virtual Reality (VR) sickness commonly known as cybersickness is one of the major problem for comfortable use of VR systems. Researchers have proposed different approaches for predicting cybersickness from bio-physiological data (e.g., heart rate, breathing rate, electroencephalogram). However, collecting bio-physiological data often requires external sensors, limiting locomotion and 3D-object manipulation during the virtual reality (VR) experience. Limited research has been done to predict cybersickness from the data readily available from the integrated sensors in head-mounted displays (HMDs) (e.g., head-tracking, eye-tracking, motion features), allowing free locomotion and 3D-object manipulation. This research proposes a novel deep fusion network to predict cybersickness severity from heterogeneous data readily available from the integrated HMD sensors. We extracted 1755 stereoscopic videos, eye-tracking, and head-tracking data along with the corresponding self-reported cybersickness severity collected from 30 participants during their VR gameplay. We applied several deep fusion approaches with the heterogeneous data collected from the participants. Our results suggest that cybersickness can be predicted with an accuracy of 87.77% and a root-mean-square error of 0.51 when using only eye-tracking and head-tracking data. We concluded that eye-tracking and head-tracking data are well suited for a standalone cybersickness prediction framework.

***Keywords*** Cybersickness Prediction, Visually induced motion sickness, Eye-tracking, Multimodal Deep Fusion Network

## 1 Introduction

Virtual reality(VR) has gained immense popularity in recent years with the rapid development of head-mounted displays (HMDs). Current HMDs can render virtual stereoscopic images at 90Hz with a resolution of 2160 x 2160 per eye, and $180°$ field of view (FOV) [1]. However, VR experience can often induce motion sickness, commonly referred to as cybersickness or visually induced motion sickness(VIMS) [2, 3]. Cybersickness-related discomforts during VR gameplay include dizziness, nausea, stomach awareness, headache, eyestrain, disorientation, fatigue, etc. [4]. These discomforts pose a significant threat to the comfortable use of VR [5, 6]. In order to reduce the effect of cybersickness,

---

\*shovonis09@gmail.com
†kevin.desai@utsa.edu
‡john.quarles@utsa.edu

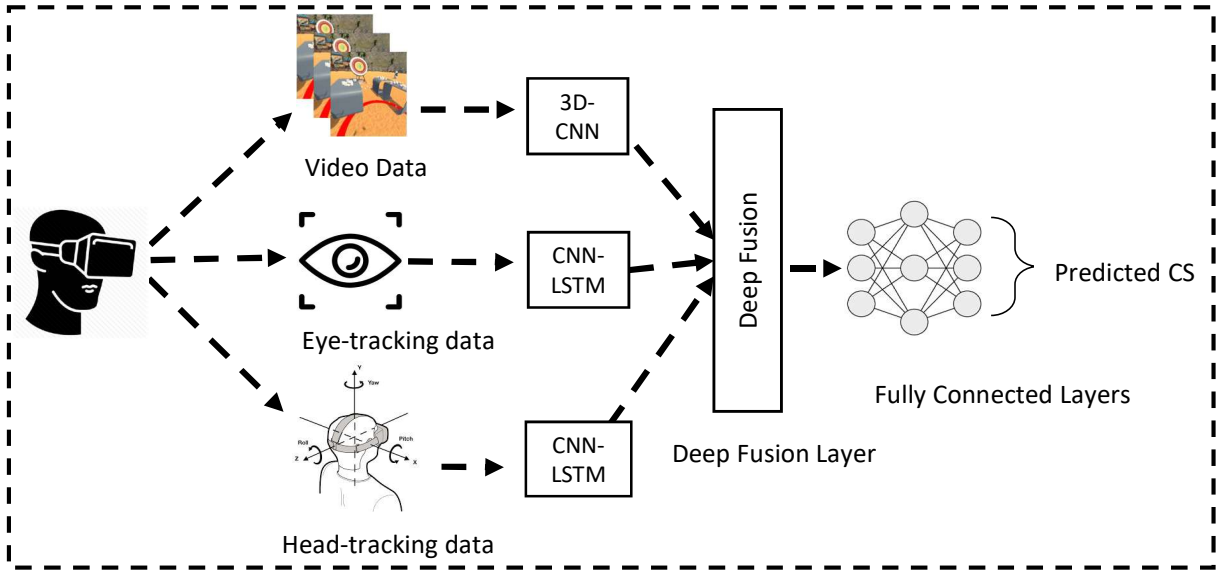

Figure 1: Overview of the proposed deep fusion approach. The model takes stereo-images, eye-tracking, and head-tracking data, which are readily available from the HMDs, and can be used in a deep fusion approach for cybersickness severity prediction

it is vital to know when the users feel cybersickness during the VR immersion. Real-time cybersickness prediction could be used in the future to automate cybersickness reduction techniques during VR gameplay [7].

Prior research had success in assessing and predicting cybersickness with bio-physiological signals (e.g., heart rate, galvanic skin responses, electroencephalogram) [8, 9, 10, 11]. However, collecting bio-physiological signals requires external sensors, which are hard to deploy as a standalone cybersickness predictor in current consumer-level HMDs (e.g., Oculus Quest, HTC Vive, Valve Index, etc.). Moreover, users are often instructed to limit their movements during the bio-physiological (e.g., electroencephalogram, heart-rate) data collection process to avoid noisy data [8, 12, 11, 10]. Thus, using external sensors can limit VR locomotion and 3D-object manipulation during the immersion, often requiring tethering and affixing sensors to the users' hands. This is not conducive to room-scale VR experiences (e.g., Beat Saber, Half-Life: Alyx). Motivated by the above-mentioned limitations, recent research is trying to predict cybersickness from the data that are readily available from the integrated sensors (e.g., eye-tracking, head-tracking) of the current consumer-level HMDs [13, 14].

In addition to using bio-physiological signals, researchers also have investigated cybersickness prediction from stereoscopic video [15, 12, 16]. In most prior research, a short pre-recorded stereoscopic video was rendered in the HMDs [15, 16]. Other research used stereoscopic images to predict visual discomforts [17, 18]. However, rendering pre-recorded stereoscopic video with an HMD does not allow VR locomotion and 3D-object manipulation during VR gameplay; thus, they are not an ideal sample of VR experiences. In addition to that, most of the stereoscopic videos used by prior research were very short (e.g., 1-2mins), which is not a representative sample as cybersickness usually grows over time [19, 20]. Therefore generalizable VR experiences should offer the participants different types of locomotion (i.e., walking, teleportation), 3D-object manipulation (i.e., touching and interacting with 3D objects), and sufficient exposure time.

In this research, we propose a novel deep fusion approach for predicting cybersickness from the data that are readily available in many current HMDs (Figure 1) (e.g., eye-tracking, head-tracking, gameplay stereoscopic video). In addition to that, the VR simulations that we used to collect our data allow the users to perform different types of locomotion and 3D-object manipulation, which are also generalizable to consumer VR games. We collected VR gameplay stereoscopic image data from 30 healthy participants along with the eye-tracking and head-tracking data. Cybersickness ground truth was constructed using a shortened version of verbally reported fast-motion scale (FMS) [21, 10] rating from the participants at each 30s interval during VR gameplay. In short our **contribution includes**:

- A novel dataset (SET) includes stereo-images, eye-tracking, and head-tracking for cybersickness research.
- A deep fusion approach to predict and classify cybersickness severity from the SET dataset.

- Insight into how eye movements significantly change due to cybersickness.
- An evaluation of the accuracy and performance of the proposed deep fusion model on SET dataset.

Our proposed deep fusion approach with eye-tracking (e.g., pupil diameter, gaze data) and head-tracking (head rotation and orientation) predicted the FMS score with a root-mean-square (RMSE) value of 0.51 on SET, which can be considered low error based on previous research on other cybersickness datasets [15, 16]. In addition, we also classified the severity of cybersickness and achieved an accuracy of 88.77% in classifying the severity of cybersickness from eye-tracking and head-tracking data.

## 2 Related Works

Cybersickness is often described as the motion-sickness-like symptoms that usually happen during or after an immersive experience. These symptoms include but are not limited to dizziness, nausea, stomach awareness, burping, etc., and can last up to a week [22]. Although cybersickness is often referred to as motion sickness for immersive systems, the root causes are different [3]. The popular theories for cybersickness are sensory conflict theory, poison theory, and postural instability theory [4, 23]. Among these theories, the sensory conflict theory is the most accepted one, which states that cybersickness occurs due to pseudo-motion-perception perceived by human visual stimuli, whereas in reality, the person is stationary. Other factors of cybersickness are display technologies, flickering, lag, and individual differences [4].

Cybersickness often causes changes in human bio-physiological signals. Researchers reported that heart rate (HR) and galvanic skin responses (GSR), eye-blink rate, pupil diameter, and electroencephalogram (EEG) change significantly due to the onset of cybersickness [9, 24, 25] and found a significant positive correlation between cybersickness and HR, EEG delta waves, and negative correlation with EEG beta-waves[9, 26]. Other researchers reported that GSR on the forehead has a higher correlation with cybersickness and could be used to predict cybersickness [27, 28]. Motivated by the significant correlation between cybersickness and bio-physiological signals, prior research proposed different approaches in predicting cybersickness using bio-physiological signals [29, 30]. Islam et al. [10] used HR and GSR to predict cybersickness with an accuracy of 87.38% using deep neural networks that were collected from 22 participants. Kim et al. [11] reported accuracy of 89.16% using EEG data. They collected EEG data from 200 participants using 8-channel EEG while the participants were immersed in 44 different VR simulations. However, most of the cybersickness studies involving bio-physiological signals are only limited to seated conditions with limited locomotion and 3D-object manipulation to avoid artifacts during bio-physiological data collection [31, 32]. In addition, using bio-physiological data requires external sensors for cybersickness prediction [11, 30]. For example, Islam et al. used NeuLog GSR sensors attached to participant's fingers, and Kim et al. used 8-channel EEG sensors, which makes it harder to deploy a closed-loop standalone cybersickness prediction and reduction system [33].

Recent research has investigated the efficacy of using stereo-image datasets collected from VR Videos to predict cybersickness [15, 16, 12]. Padmanaban et al. used 19 VR videos (i.e., 2 mins length) as their dataset and used depth and optical flow features to predict cybersickness with a root-mean-square error (RMSE) of 12.6. Later, Lee et al. improved the results with an RMSE value of 8.49 by using a 3D-convolutional neural network [34] and using a multimodal deep fusion approach with optical-flow, disparity, and saliency features. Kim et al. [12], used a convolutional auto-encoder [35] to predict cybersickness by utilizing reconstruction error captured from exceptional motion videos (i.e., fast motion video). However, the videos used in most of the prior research are pre-recorded and were rendered using the HMDs, instead of allowing the participants to interact in the VR simulations. For example, Kim et al. [12] used KITTI dataset [36], which are not VR videos. Padmanaban and Lee et al. used pre-recorded short VR videos (1-2 mins long), which also did not allow free locomotion and 3D-object manipulation and may not have included a long enough exposure to introduce cybersickness [19, 20]. Jin et al. [37] used five different VR videos and allowed users to do different types of locomotion and 3D-object manipulation and achieved a coefficient of determination ($R^2$) value of 86.8% in predicting cybersickness from the video features.

Due to the limitations of the approaches mentioned above of cybersickness prediction from stereoscopic video and bio-physiological signals, recent research has been focused on predicting cybersickness from eye-tracking, and motion data [13, 38, 39, 40]. Chang et al. [13] reported that different eye features (e.g., fixation duration and distance between the eye gaze and the object position sequence) are highly correlated when the participants felt cybersickness and proposed and support vector machine (SVM) regression for cybersickness prediction. Lopes et al. [38] reported that pupil position and eye-blink rate between the sickness group and the non-sickness groups were significantly different. Guo et al. examined the use of optokinetic after nystagmus (OKAN) to predict cybersickness reported that the cybersickness group had a consistent pattern of OKAN, which had a significant correlation with the corresponding subjective measures. Some other researchers also investigated using head-tracking and postural data for cybersickness prediction [14, 19, 41, 40]. Feigl et al. used motion parameter as an objective measure of cybersickness [40]. Other research reported that postural instability when standing can be used to predict the likelihood of cybersickness [19].

In addition to using objective measurements (i.e., bio-physiological signals, stereoscopic video) for cybersickness studies, researchers often use subjective measures to detect cybersickness severity. The most commonly used cybersickness subjective measures are simulator sickness questionnaire (SSQ) [22]. However, several researchers have argued that cybersickness is different from simulator sickness [3, 42] and proposed cybersickness susceptibility questionnaire [43] and virtual reality sickness questionnaire [44] for subjective measurement of cybersickness. However, these subjective measures are often collected after the VR immersion. Therefore they do not provide sufficient granular understanding of cybersickness severity during VR immersion. Motivated by the shortcomings of subjective measures after VR immersion, Keshavarz et al. proposed the Fast Motion Scale (FMS) [21] to collect subjective measures during VR immersion. In addition, jin et al. [37] used a Ranking-Rating (RR) method to rate cybersickness during VR immersion on a scale from [0-10]. They reported a significant correlation (i.e., .838) with the RR method and SSQ, which they used for their ground-truth construction. Similar to Jin et al., and [37], and Keshavarz et al. [21] we used a shortened $FMS$ scale [0-10] to collect the subjective measures of cybersickness during VR immersion.

This research presents a novel approach for cybersickness prediction from the readily available HMD's integrated sensors. More specifically, we used eye-tracking, head-tracking, and stereoscopic video data for cybersickness prediction. In addition to that, we also used five different VR experiences that allowed the participants different locomotion and 3D-object manipulations. We had significant prediction accuracy by using a deep fusion approach with eye and head-tracking data, which we believe can be used to develop a standalone cybersickness prediction framework. We plan to explore this direction in future studies.

## 3 User Study

Motivated by the limitations of the currently available cybersickness datasets that include eye-tracking, head-tracking, and stereo-images, we conducted our data collection through a user study. We proposed a total of 1755 short stereoscopic videos and their corresponding eye-tracking and head-tracking data, collected from the 30 participants (i.e., SET dataset). The data were extracted from five different VR simulations. The following subsection details the user study and the proposed SET dataset.

### 3.1 Virtual Environments

We used a total of five VR simulations in our study (Figure 2). Each simulation lasted 7 minutes long unless the participants choose to terminate earlier. The study simulations were developed based on four criteria: a) Locomotion, b) Use of Controllers, c) Experience, and d) Motion Perception. The simulations are as follows:

*Beach City:* In the *Beach City* simulation, participants used both the controllers to navigate and manipulate 3D objects. The simulation was rendered at room scale and was experienced while standing. The participants were allowed to do both teleportation and natural walking. The task was to shoot the virtual targets with the bow and arrow from different pre-defined locations in the virtual environment. Participants used both controllers for 3D-object manipulation.

*Road Side:* The task for the *Road Side* simulation was similar to the *Beach City* simulation. However, the participants were only allowed to navigate by walking, and teleportation was disabled. Participants walked in the virtual environment to different pre-defined locations to shoot the flying balloons and the targets using the longbow and arrow. The simulation was experienced at room scale while standing, and participants were allowed to use both the controllers.

*Furniture Shop:* In the *Furniture Shop* simulation, participants were only allowed to do teleportation for navigation. The task was to find the longbow hidden in a virtual room environment and shoot the virtual targets with the longbow. Participants were also allowed to interact with the 3D objects in the virtual environment. The experience was room-scale, and participants could use both controllers. Participants were also allowed to experience the simulation in both "Seated" and "Room-scale" conditions.

*SeaVoyage:* In the *SeaVoyage* simulation, participants were placed in a virtual boat, and the task was to steer the boat using one controller (i.e., participants were allowed to choose) and pass all the checkpoints around the island. Participants experienced the *SeaVoyage* simulation in a seated condition, and the participants controlled the motion of the boat.

*Roller Coaster:* Participants experienced the *Roller coaster* in a seated condition. The roller coaster takes 57s to complete a cycle, and there was a pause of 10s between two consecutive roller coaster cycles. Participants could freely orient their view but did not control the movement of the roller coaster, and none of the controllers were used.

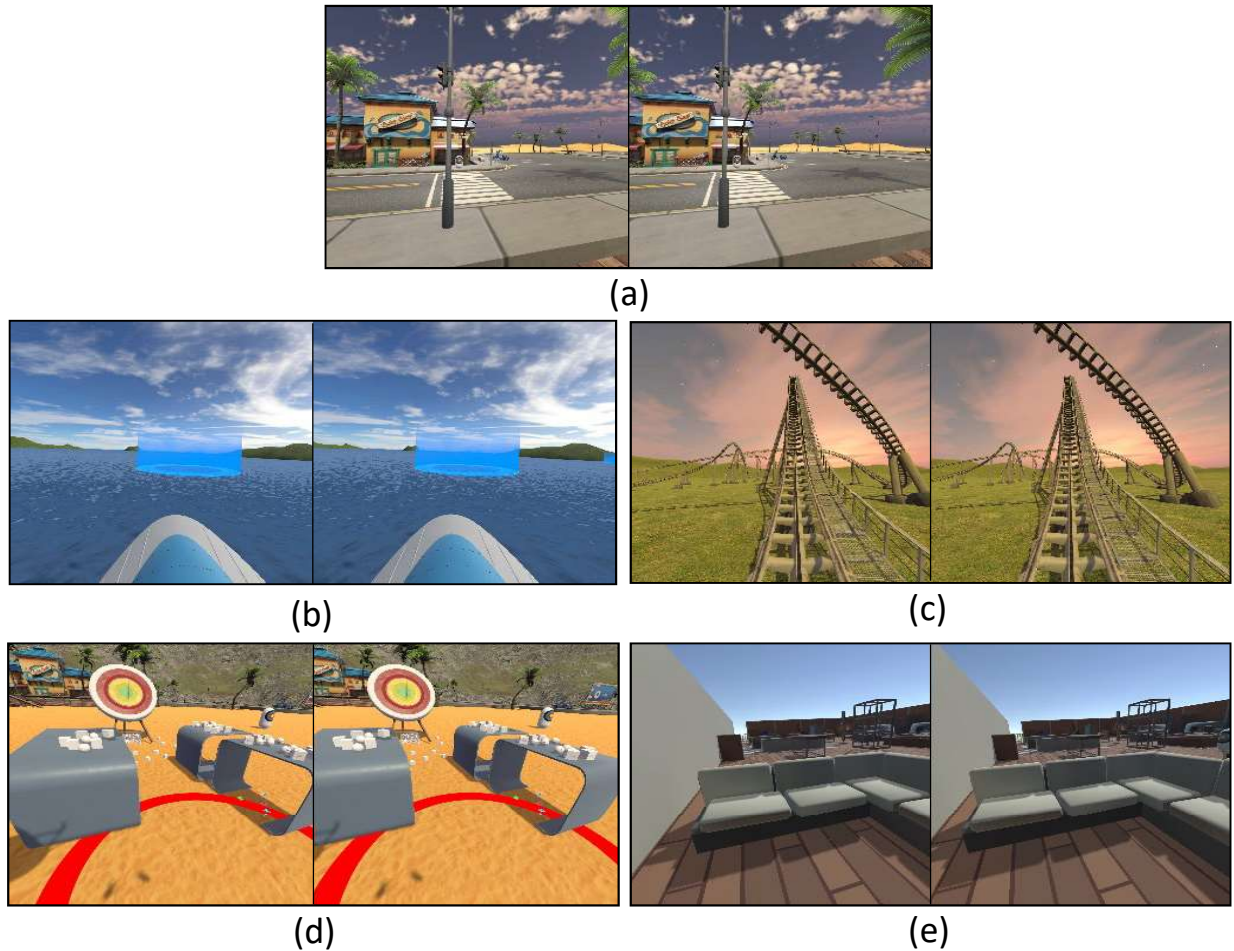

Figure 2: VR simulations used for the study. (a) Road Side Simulation, (b) The SeaVoyage Simulation, (c) VR Roller Coaster Simulation, (d)Beach City Simulation, (e) Furniture Shop Simulation

## 3.2 Apparatus

We used an HTC-Vive Pro Eye for rendering the virtual simulations with a resolution of 1440 x 1600 per eye and a refresh rate of 90Hz. The field of view of the HMD was 110°. The sound was played from the integrated Vive headphones. The HMD was also configured with the Vive wireless adapter to allow the participants a smooth VR experience. All the simulations were executed on an Intel i7 CPU with 32Gb of memory and Nvidia RTX 2070 GPU. The virtual reality simulations were developed using Unity 3D. The *Dynamic Water Physics 2* [4] asset was used for the water simulations, and *Beach City Props* [5] asset was used for the *Beach City* simulation. For collecting the eye-tracking data (e.g., gaze, pupil diameter), we used HTC SRanipal SDK and Tobii HTC Vive Devkit. In order to collect all the data safely without blocking the VR simulation, we developed an asynchronous LogAPI module [6] for Unity software. Table 2 summarizes the raw data collected from using the LogAPI, which we plan to make open-source in the future. In order to sanitize the HMD and the controllers, we used Cleanbox[7], which emits UVC light provides safe, hospital-grade hygiene, killing 99.99% of contagions.

---

[4]https://bit.ly/3bX7D6n
[5]https://bit.ly/3eZLDtm
[6]https://github.com/shovonis/HMDLogManager
[7]https://cleanboxtech.com

Table 1: Virtual Reality Simulations with different conditions used in this study

| Simulation Name | Locomotion | | Use of Controllers | | Experience | | Motion Perception | |
|:---:|:---:|:---:|:---:|:---:|:---:|:---:|:---:|:---:|
| | *Walking* | *Teleportation* | *Left* | *Right* | *Seated* | *Room Scale* | *Self Controlled* | *Simulated* |
| *Beach City* | ✓ | ✓ | ✓ | ✓ | ✗ | ✓ | ✓ | ✗ |
| *Road Side* | ✓ | ✗ | ✓ | ✓ | ✗ | ✓ | ✓ | ✗ |
| *Furniture Shop* | ✗ | ✓ | ✓ | ✓ | ✓ | ✓ | ✓ | ✗ |
| *Sea Voyage* | ✗ | ✗ | ✓ | ✗ | ✓ | ✗ | ✓ | ✗ |
| *Roller Coaster* | ✗ | ✗ | ✗ | ✗ | ✓ | ✗ | ✗ | ✓ |

### 3.3 Participants

We recruited 30 participants (M:15, F15) for our user study. The mean age was 29.04 years, and the standard deviation was 9.26 years. Their races include *Asian, Black Caucasian, Hispanic and Pacific Islander*. None of the participants reported any vestibular dysfunction and were not taking any medication related to motion sickness. All the participants had normal or corrected-to-normal vision. The data were safely collected using an Institutional Review Board (IRB)-approved COVID-19 protocol reviewed by a medical professional. All the participants filled out a verbal and written COVID-19 screening questionnaire before starting the study. The temperature reading on the forehead was also taken before entering the lab facility. Participants were instructed to wear a facial mask during the study, and a 6ft distance between the participants and the experimenters was ensured. Among the 30 participants, three participants' data could not be used due to technical issues (i.e., Bluetooth and HTC-Vive wireless adapter black screen issue[8]). Therefore for this study, we only considered the data collected from 27 participants. All the participants were paid $30 hourly for their time and efforts, and each study took approximately 2 hours.

### 3.4 Data Collection Procedure

Participants were instructed to verbally report their discomfort on an FMS scale [21] from [0-10] whenever a pre-recorded voice prompt was played at each 30s interval during the VR simulations. Rating "0" means no significant change of discomfort compared to resting baseline, and rating "10" means significant discomfort compared to resting conditions. Participants had the flexibility to quit or postpone the study at any point in time. Data were collected using the following steps:

### 3.5 Data Collection Procedure

Participants were instructed to verbally report their discomfort on an FMS scale [21] from [0-10] whenever a pre-recorded voice prompt was played at each 30s interval during the VR simulations. Rating "0" means no significant change of discomfort compared to resting baseline, and rating "10" means significant discomfort compared to resting conditions. Participants had the flexibility to quit or postpone the study at any point in time. Data were collected using the following steps:

1. First, the participants filled out their demographic information and whether they have a prior immersive VR experience.
2. Participants then completed a training VR session to understand how to use the controllers for different locomotion techniques(e.g., teleportation) and 3D object manipulation.
3. After completing the VR training session, the five VR simulations were played in counter-balanced order. Eye and head position was calibrated before the start of each simulation. Each VR simulation lasted 7 minutes long unless the participants decided to quit earlier.
4. After completing each VR simulation, participants filled out an SSQ questionnaire and rested (i.e., participants were allowed to rest until they felt normal) before starting the next simulation.

## 4 Data Processing and Analysis

The HTC-Vive Pro Eye rendered the simulation at 90Hz. However, collecting stereo-image data at 90Hz per frame is a computationally expensive process that caused flickering to the simulation (i.e., I/O operations to save 90 images per frame refresh causes flickering to the VR simulation). Therefore, all the data were collected with a 20Hz sampling rate to minimize flickering issues. We extracted a total of 9 features from the eye-tracking data (i.e., pupil diameter of both eyes, normalized gaze direction, convergence distance) and four features from the Head-Tracking data (i.e., Head quaternion values-x, y, z, and w). The eye-tracking and head-tracking data were normalized using the following formula:

$$D_i = \frac{D_i - \mu}{\sigma} \qquad (1)$$

---
[8]https://bit.ly/2RBjjVe

Table 2: List of raw data collected from the HTC-Vive Pro Eye's integrated Sensors

| Data Type | Data |
|---|---|
| Eye-Tracking Data | Left pupil diameter (mm) <br> Right pupil diameter (mm) <br> Left normalized gaze direction <br> Right normalized gaze direction <br> Convergence distance (mm) |
| Head-Tracking Data | Head Quaternion <br> Rotation (i.e., x, y, z and w) |
| Stereo Images data | Left Image <br> Right Image <br> Optical Flow <br> Disparity Map |

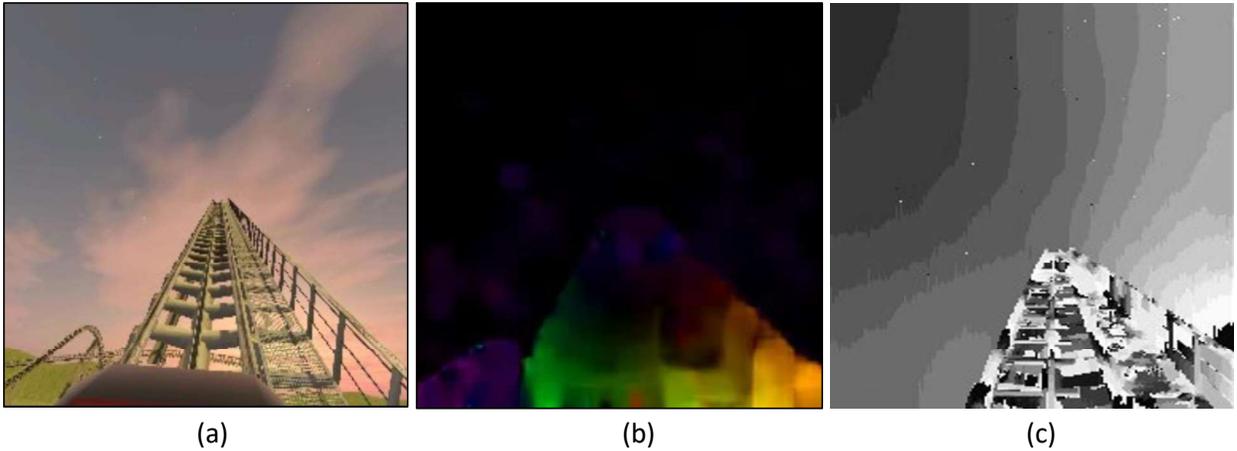

(a)　　　　　　　　　　　(b)　　　　　　　　　　　(c)

Figure 3: (a) Left view image of a stereoscopic frame from the VR roller coaster simulation, (b) The corresponding optical flow of the same frame and (c) The corresponding disparity map using the semi-global block matching method.

Here, $\mu$ represents the mean of the data, and $\sigma$ is the standard deviation of the data. For stereo images, we extracted the left and right eye images at 256 x 256 resolution. Table 2 summarizes the collected raw features from the integrated HMD's sensors.

We used only the left view image for optical flow calculation similar to prior works [15, 16]. The optical flow was calculated using the dense optical flow algorithm proposed by Gunnar Farneback [45]. The motion displacement was represented using the RGB color representation, and the optical flow was transformed to the polar HSV values. Finally, the calculated optical flow is mapped back to RGB representation (Figure 3.b).

A disparity map represents the horizontal differences of each pixel between the left and right view of the stereo images, which prior researchers used to predict cybersickness [15, 16]. Similar to Lee et al., [16] we used a semi-global block matching [46] method with a filter of size (5 x 5) to obtain our disparity maps from the stereo images (Figure 3.c)

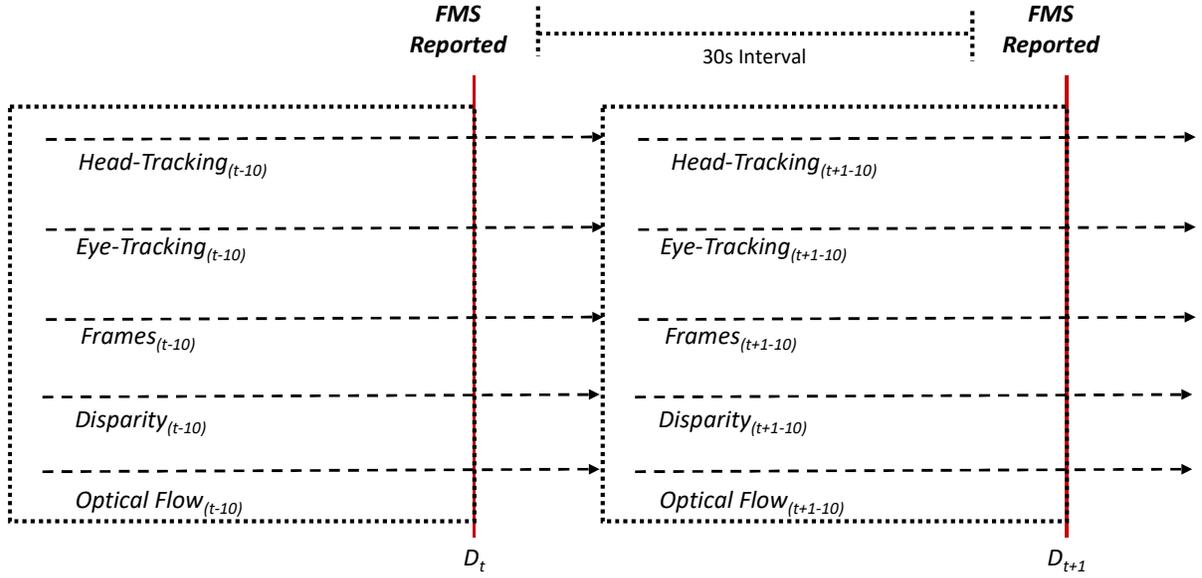

Figure 4: Data slicing process and ground truth construction. $D_t$ represents one data point at time $t$, which consists of the optical flow, disparity, frames, eye-tracking and head-tracking data from $(t-10)_s$ to $t_s$.

Table 3: Summary of the paired t-test of the gaze data (normalized-X) between the sickness and non-sickness measures (df=26). * denotes significant with **p-value** $< .05$

| Simulations | Mean | SD | t-value |
|---:|---:|---:|---:|
| Beach City (Non-Sickness)* | -0.012 | 0.13 | |
| Beach City (Sickness) | -0.043 | 0.15 | 3.52 |
| Road Side(Non-Sickness)* | -0.0031 | 0.146 | |
| Road Side(Sickness) | 0.028 | 0.152 | -3.28 |
| Furniture Shop (Non-Sickness) | -0.007 | 0.118 | |
| Furniture Shop (Sickness) | -0.008 | 0.134 | -0.014 |
| Sea Voyage (Non-Sickness)* | -0.013 | 0.15 | |
| Sea Voyage (Sickness ) | 0.009 | .11 | -2.55 |
| Roller Coaster (Non-Sickness)* | -0.013 | 0.13 | |
| Roller Coaster (Sickness ) | 0.004 | 0.10 | -2.30 |

## 4.1 Ground-Truth Construction

In order to construct the ground truth, we used the FMS score collected at each 30s interval during the VR simulations. Although we have collected SSQ after each simulation, SSQ scores do not provide granular cybersickness severity during the simulation; therefore, we used the FMS scores for ground truth construction in this study.

### 4.1.1 Cybersickness Regression

Let's consider that at time $t$ the user reported their current $FMS_t$ on a scale from [0-10]. Since cybersickness provokes delay in reaction time [24, 47], the point in time when the participants reported $FMS_t$ might not be the exact time when they started feeling that corresponding discomforts. Therefore, we considered all the data, D (i.e., frames, eye-tracking and head-tracking) from $(t-10)_s$ to $t_s$ to construct the region of interests data points associated with the corresponding $FMS_t$ (Figure 4). Formally, the ground-truth can be represented as follow:

$$[D^{t-10}, D^{t-9}, D^{t-8}, ..., D^t] \to FMS_t$$

Here, D represents the consecutive frames, optical flow, disparity map, eye-tracking, and head-tracking data sampled at 20Hz. We collected a total of 11s (i.e., including data $D_t$ ) of data. The task of regression is to predict the $FMS_t$ given the data from $[D_{(t-10)} - D_t]$.

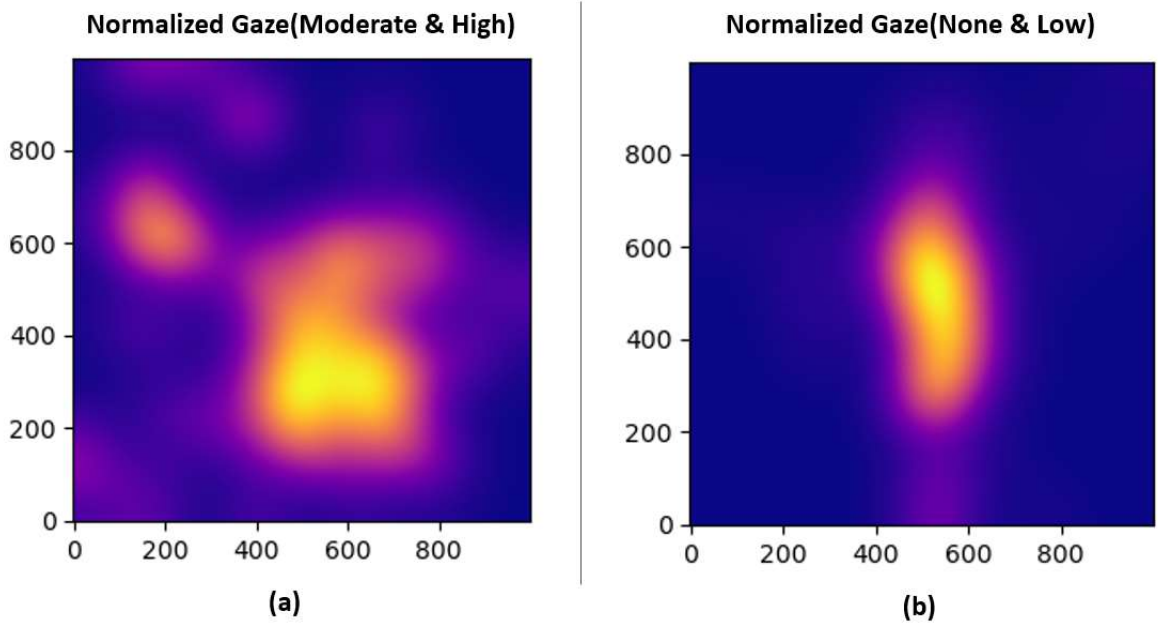

Figure 5: Normalized Gaze Heat Map of a particular Individual for *"Beach City"* Simulation (a) Normalized gaze heat-map when the Individual felt moderate to high sickness (b) Normalized gaze heat-map when the Individual felt None or low sickness.

Table 4: Summary of the paired t-test of the gaze data (normalized-Y) between the sickness and non-sickness measures (df=26). * denotes significant with **p-value** $< .05$

| Simulations | Mean | SD | t-value |
|---:|---:|---:|---:|
| Beach City (Non-Sickness)* | -0.114 | 0.132 | |
| Beach City (Sickness) | -0.130 | 0.110 | 2.01 |
| Road Side(Non-Sickness)* | -0.114 | 0.140 | |
| Road Side(Sickness) | -0.158 | 0.117 | 5.36 |
| Furniture Shop (Non-Sickness) | -0.053 | 0.130 | |
| Furniture Shop (Sickness) | -0.056 | 0.133 | 0.35 |
| Sea Voyage (Non-Sickness)* | -0.03 | 0.096 | |
| Sea Voyage (Sickness ) | 0.051 | 0.080 | -14.93 |
| Roller Coaster (Non-Sickness)* | -0.175 | 0.159 | |
| Roller Coaster (Sickness ) | -0.099 | 0.172 | -7.26 |

### 4.1.2 Cybersickness Severity Classification

We conducted a distribution analysis on the collected FMS scores from all participants for the cybersickness severity classification task similar to prior research [10]. The first quantile ($Q_1$) of the distribution was 1.0, the second quantile ($Q_2$) of the distribution was 2.0, and the third quantile ($Q_3$) of the distribution was 4.0. We used the following rule for cybersickness severity classification $CS_t$ similar to a prior study [10].

$$CS_t = \begin{cases} None, & \text{if, } 0 \leq FMS_t \leq Q_1 \\ Low, & \text{if, } Q_1 < FMS_t \leq Q_2 \\ Medium, & \text{if, } Q_2 < FMS_t \leq Q_3 \\ High, & \text{if, } FMS_t > Q_3 \end{cases} \quad (2)$$

Based on the $FMS_t$ the cybersickness severity ($CS_t$) was categorized into four groups: 1) None, 2) Low, 3) Medium and 4) High. The classification task is to classify $CS_t$ given the data points $[D_{(t-10)} - D_t]$.

Table 5: Summary of the paired t-test of the pupil diameter (mm) between the sickness and non-sickness measures (df=26). * denotes significant with **p-value** $< .05$

| Simulations | Mean | SD | t-value |
| --- | --- | --- | --- |
| Beach City (Non-Sickness)* | 3.19 | 1.30 | |
| Beach City (Sickness) | 2.99 | 0.86 | 3.01 |
| Road Side (Non-Sickness)* | 3.20 | 1.16 | |
| Road Side (Sickness) | 2.79 | 0.76 | 5.56 |
| Furniture Shop (Non-Sickness) | 2.50 | 1.64 | |
| Furniture Shop (Sickness) | 2.55 | 1.60 | -0.34 |
| Sea Voyage (Non-Sickness)* | 2.82 | 0.35 | |
| Sea Voyage (Sickness) | 2.27 | 1.01 | 11.55 |
| Roller Coaster (Non-Sickness)* | 3.56 | 0.83 | |
| Roller Coaster (Sickness) | 3.00 | 1.01 | 9.29 |

We had a total of 27 participants and five VR simulations, each lasting 7 minutes. The FMS was collected at each 30s interval. We extracted 1755 stereoscopic videos with their corresponding eye-tracking and head-tracking data. Each video was 11s seconds long; therefore, we had a total data sample of 19,305. (i.e., each $D_t$ contains 11s of stereoscopic video, corresponding eye-tracking, and head-tracking data).

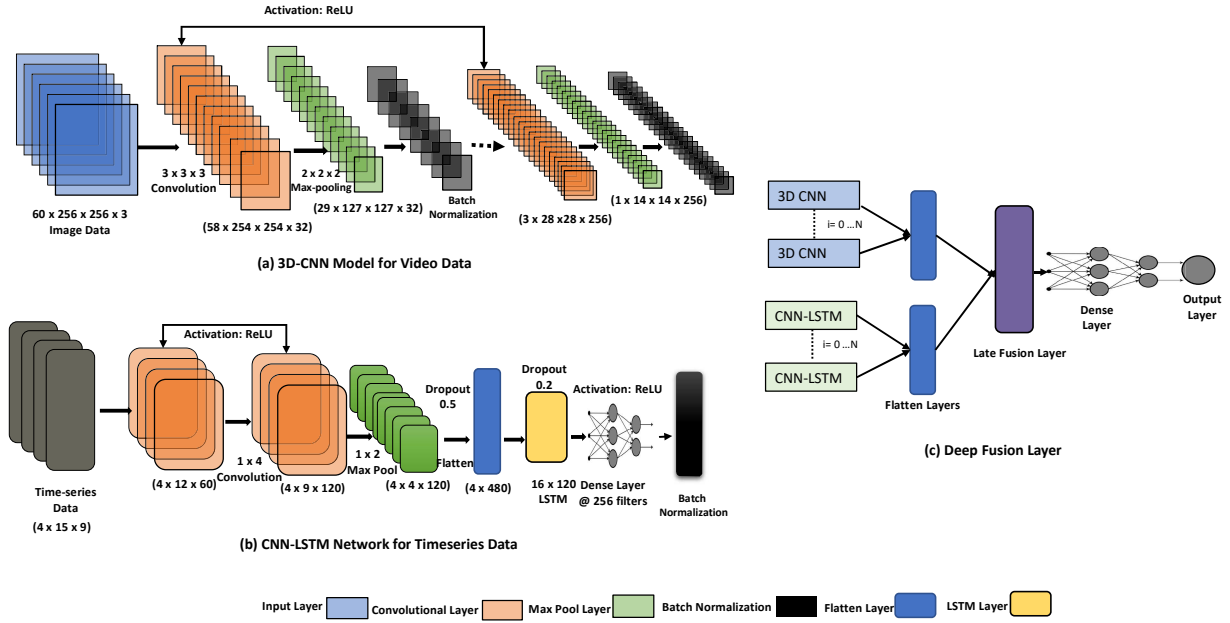

Figure 6: Overview of the proposed Deep Fusion Model. (a) Proposed 3D-CNN architecture for the stereoscopic video data, (b) Proposed Time Distributed CNN-LSTM layer for Eye-tracking and Head-Tracking data and (c) The Deep Fusion Layer to fuse different modalities and predict cybersickness

## 4.2 Data Analysis

We conducted an initial data analysis of the eye-tracking and head-tracking data. Figure 5 illustrated the normalized gaze heat-map when a participant reported they felt moderate to severe cybersickness (Figure 5.a) and when the same participant reported they were not feeling cybersickness (Figure 5.b). The figure shows that the participant fixated their gaze in the middle of the screen without any cybersickness and gazed all over the screen when feeling cybersickness.

We also conducted paired t-tests of the eye-tracking and head-tracking data between the non-sickness (i.e., When the participant's $FMS_t \leq 2$) and sickness (i.e., when same participant's $FMS_t > 2$)). For these paired t-tests, we used the repeated measure data from all 27 participants, grouping and averaging their eye-tracking data into sickness and non-sickness measurements for each VR simulation. Each simulation had a different task and different types of locomotion and motion perception. So the eye-tracking data from one simulation was not directly comparable with other simulations; therefore, we report the paired t-test analysis for each simulation. We have found significant differences (p < .05) in the gaze data and pupil-diameter between sickness and non-sickness measurements for all the simulations except the "Furniture Shop" simulation. Paired t-test results for the "Gaze data" for all the simulations are summarized in Table 3 and 4 . Significant differences(p < .05) in pupil diameter were also found for the simulations (Table 5) between sickness and non-sickness measurements. During the "Furniture Shop" Simulation, only two participants reported $FMS_t > 2$, and no significant difference(p > .05) in gaze data and pupil diameter were found. For the head-tracking data, we did not find any significant differences in the quaternion values (x, y, z, and w) of head rotation grouped by the sickness and the non-sickness data from each simulation ($p > .05$).

Table 6: Mean Accuracy, Precision and Recall of the 10-Fold Cross Validation on Cybersickness Severity Classification

| Fusing Modalities | % Accuracy | % Precision | | | | % Recall | | | |
|---|---|---|---|---|---|---|---|---|---|
| | | *None* | *Low* | *Medium* | *High* | *None* | *Low* | *Medium* | *High* |
| *Video* | 45.83 | 66.67 | 50.0 | 83.33 | 62.50 | 1.0 | 25.0 | 55.55 | 71.42 |
| *Video + Optical Flow* | 54.17 | 87.50 | 42.85 | 80.0 | 87.50 | 87.50 | 75.0 | 44.44 | 98.56 |
| *Video + Opt. + Disparity* | 50.78 | 92.05 | 50.0 | 85.71 | 87.5 | 87.50 | 75.0 | 66.67 | 98.32 |
| *Video + Opt. + Disp. + Eye + Head* | 52.15 | 95.35 | 54.41 | 62.71 | 53.84 | 50.21 | 55.55 | 66.67 | 94.38 |
| *Head-Tracking* | 73.2 | 73.94 | 66.82 | 73.57 | 79.79 | 88.42 | 81.03 | 41.80 | 81.63 |
| *Eye-Tracking* | 80.7 | 87.31 | 72.41 | 75.88 | 85.90 | 98.11 | 68.33 | 72.12 | 84.37 |
| **Eye + Head Tracking** | **87.7** | **96.78** | **83.34** | **79.58** | **91.73** | **96.23** | **78.87** | **85.87** | **89.92** |

## 5 Multi-modal Neural Network

We proposed a multi-modal deep fusion network for predicting the $FMS_t$ and classifying $CS_t$. The model takes input from the heterogeneous data inputs $D_t$. We used a 3D-CNN[34] model for the video data. In order to learn features from the eye-tracking and the head-tracking data, we used a Time-distributed CNN-LSTM model similar to Islam et al., [10]. Features learned from each modalities are fused together using a late fusion approach [48] (Figure 6).

### 5.1 3D-CNN For Learning Spatiotemporal Features

We used multi-layered 3D-CNN architecture for learning the spatiotemporal features from the stereoscopic video, optical flow, and disparity map (Figure 6). The model takes input as a sequence of consecutive frames from the video source and learns the hidden spatiotemporal features. The timestep is set to 60, which means that the model will take 60 consecutive frames at a time at a single input. Therefore, the input shape of the model is (60 x 256 x 256 x 3). The kernel size for the convolutional layer is set to (3 x 3 x 3). We used ReLU activation functions[49] for the convolutional layer, and the L2 kernel regularizes with the regularization factor as 0.01. A 3D max-pool layer follows the 3D-convolutional layer with a pool size of (2 x 2 x 2) with a stride of 2. After the 3D max-pool layer, we performed a batch-normalization to reduce batch overfitting. We used a total of 3 sets of convolutional, max-pool, and batch-normalization operations. The final output is flattened, which was used as input for the late-fusion operation. A summary of the 3D-CNN operations is illustrated in Figure 6 (a).

### 5.2 Time-Distributed CNN-LSTM

We used Time-distributed CNN-LSTM layers for learning the features from the eye-tracking and head-tracking data (Figure 6.b). The timestep is set to 60, forwarded into four subsequence time-distributed 1-D convolutional layers of shape (4 x 15). This process allows the network to learn the four subsequences of features simultaneously from the eye and head-tracking data. The eye-tracking data have nine features, and the head-tracking data have four features (Table 2). Therefore, the eye-tracking input data shape is (4 x 15 x 9), and the head-tracking input data shape is (4 x 15 x 4). We used ReLU as the activation function for the convolutional layers. The convolution layers are followed by a time-distributed 1D max-pool layer with a stride of size two and a pool size of 2. We add a Dropout layer with a dropout rate of 50% to reduce the overfitting of the model. The output from the dropout layer is then flattened for the next LSTM layer, which learns the temporal features from the data. The LSTM layer had a recurrent dropout of 20%, and the output is followed by a fully connected Dense layer with 256 filters. The output from the Dense layer is then flattened and forwarded to a batch-normalization layer to reduce batch overfitting of the data.

Table 7: Mean $R^2$, $PLCC$ and $RMSE$ of the 10-fold Cross Validation of the FMS Regression

| Fusing Modalities | $R^2$ | $PLCC$ | $RMSE$ |
|---|---|---|---|
| *Video* | 0.12 | 0.27 | 4.23 |
| *Video + Optical Flow* | 0.15 | 0.34 | 3.89 |
| *Video + Optical + Disparity* | 0.09 | 0.24 | 3.65 |
| *Video + Opt. + Disp. + Eye + Head* | 0.02 | 0.25 | 3.27 |
| *Head-Tracking* | 0.18 | 0.52 | 1.23 |
| *Eye-Tracking* | 0.56 | 0.79 | 0.93 |
| ***Eye + Head Tracking*** | **0.67** | **0.84** | **0.51** |

### 5.3 Deep Fusion Layer

The flattened outputs from the 3D-CNN and the CNN-LSTM layers are fused together using a concatenate layer. The output from the concatenate layer is then passed to a fully connected Dense Layer with 256 neurons and ReLU as the activation function (Figure 6 (c)). The final output layer then outputs the result. For cybersickness severity classification, we used 'Softmax' as the activation function with four neurons. The softmax function outputs the probability of each one of the classes (e.g., None, Low, Medium, or High), and the maximum probability value is used to predict the cybersickness severity class. We used several late fusion approaches from the different modalities in Table 2. For cybersickness $FMS_t$ prediction, we used the Linear activation function for the final Dense output layer with one neuron, which outputs the predicted $\hat{FMS}_t$.

### 5.4 Loss Function

We used different loss functions for the regression and the classification task. For the regression task (e.g., predicting $FMS_t$) we used root-mean-square error as the loss function (Equation 3).

$$RMSE = \sqrt{\frac{\sum_{i=1}^{N}(FMS_t^i - \hat{FMS}_t^i)^2}{N}} \quad (3)$$

Here, $FMS_t^i$ is the actual verbally reported FMS score from the participant at time $t$ and $\hat{FMS}_t^i$ is the predicted FMS score by the network. N represents the number of samples. For the cybersickness severity classification task $CS_t$ we use categorical cross-entropy loss function (Equation 4).

$$L = -\sum_{i=0}^{K}(CS_t * \log(\hat{CS}_t)) \quad (4)$$

Here, $CS_t$ refers to the ground truth class of the cybersickness severity and $\hat{CS}_t$ refers to the predicted severity class, and $K$ is the total number of classes.

## 6 Experiment Setup

We used TensorFlow for training and evaluating our models. All the models were run in an NVidia DGX-1 Server with Ubuntu 18.0 operating system. The server had 2 x Intel Xeon processors and 4 x NVIDIA Tesla P100 GPU. The system had a memory of 512GB.

### 6.1 Model Evaluation

We used a 10-fold cross-validation method to train and test the performance of the proposed model similar to prior works [50, 51, 52, 15].In k-fold cross-validation, the dataset is partitioned into k groups. Out of the k partition, a single partition is used for testing the model, and the rest (k-1) partitions are used for training the model [53]. The process is repeated k times each time by selecting a different test partition and the remaining (k-1) partition as a training dataset, thus reducing any bias from the dataset [51].

Table 8: A result comparison analysis between our proposed approach using eye-tracking and head-tracking and prior approaches

| Paper Title | Accuracy | PLCC | RMSE |
|---|---|---|---|
| *Padmanaban et.al* [15] | 51.94 | -0.0078 | 12.69 |
| *Chang et.al* [13] | 34.8 | - | - |
| *Dennison et.al* [29] | 78.0 | - | - |
| *Kim et. al. [56]* | 63.12 | - | - |
| *Kim et. al. [11]* | **89.16** | - | - |
| *Proposed Method* | 87.77 | 0.8442 | 0.51 |

### 6.1.1 Hyper-Parameter and Validation

In order to fine-tune the model parameters during the training process, we used 20% of the training dataset as validation data during each fold [54].[9] We used Adam as the optimizer for our model with epochs of 300 and a batch size of 512. The parameters are set using hyper-parameter tuning. In order to prevent the model from overfitting, we deployed an early-stopping strategy with a patience value of 20 while training the model [55].

## 7 Results

The proposed multimodal deep fusion approach for predicting $FMS_t$ and classifying the cybersickness severity achieved significant accuracy while using eye-tracking and head-tracking data.

### 7.1 Cybersickness Severity Classification

The mean accuracy, precision and recall of the during the 10-fold cross validation are presented in Table 6. The proposed deep fusion approach with eye-tracking and head-tracking data achieved an accuracy of 87.7%. The model achieved higher precision and recall value for the "None" severity class, which was 96.78% and 96.23% respectively, and 91.23% and 89.91% for the "High" cybersickness severity class, respectively. However, the model struggled in differentiating between the "low" and "Moderate" classes. When only using the eye-tracking data, the model achieved an accuracy of 80.70%. The model also achieved an accuracy of 73.2% while only using head-tracking data. The precision and recall value for the eye-tracking and head-tracking data was significantly higher than the other fusion approaches. However, the accuracy of using the stereoscopic video data and other modalities resulted in poor performance. For example, while using stereoscopic video, optical flow, disparity, eye, and head-tracking data, the model only achieved an accuracy of 52.15%.

### 7.2 Cybersickness $FMS_t$ Prediction

The R-squared ($R^2$), Pearson-Linear-Correlation coefficient (PLCC) and the RMSE value for the 10-fold cross-validation for the cybersickness $FMS_t$ prediction results are summarized in Table 7. The PLCC represents the linear correlation between the predicted $\hat{FMS}_t$ and the actual $FMS_t$, while the $R^2$ value represents the proportion of the variance between them. Similar to the cybersickness classification task, the model achieved significantly better results while using eye-tracking and head-tracking data. The PLCC, RMSE, and $R^2$ value, while using eye-tracking and head-tracking data for the model, was 0.84, 0.51, and 0.67, respectively. With eye-tracking data, the model achieved a PLCC of 0.79 with an RMSE value of 0.93. Compared to prior works [15, 16], this is a significant improvement in terms of PLCC and RMSE. Although the proposed deep-fusion approach achieved significant results with eye and head-tracking data, the model struggled with fusing the video data (Table 7).

## 8 Discussion

In this research, we present a novel deep fusion approach of cybersickness prediction and classification using the data that are readily available from the integrated HMD sensors. The proposed deep fusion model achieved an accuracy of 87.7% in classifying cybersickness severity from eye-tracking and head-tracking data. The precision and recall percentage for different cybersickness severity classes were also significantly higher with fused eye-tracking and head-tracking data. The model also predicted the $FMS_t$ with an RMSE value of 0.51. The model also had significantly better precision and recall value in predicting No cybersickness and High Cybersickness (Table 6).

---

[9]Validation data is not test data and 20% validation data is randomly extracted during each fold of the 10-fold cross-validation process.

Since there is a minimal eye-tracking dataset for the cybersickness study, our results are not directly comparable with the prior works. However, we compared our results in terms of using integrated sensors versus using external sensors. Chang et al.; reported an $R^2$ value of 34.8% using integrated eye-tracking data from FOVE VR headset for cybersickness prediction [13]. Wibirama et. al reported an $R^2$ value of 4.2% using eye-tracking data [57]. Our proposed deep fusion approach with eye-tracking data only achieved an $R^2$ value of 56%, using both eye-tracking and head-tracking data; the model achieved an $R^2$ value of 67%. When using external physiological sensors, Dennison et al.; reported accuracy of 78% in classifying cybersickness severity [29]. Using the cognitive features of EEG, Kim et al. reported accuracy of 89.16% [11]. Padmanaban et al. [15] reported accuracy of 51.94% using video and optical features of cybersickness. Using a 3D-CNN architecture, Lee et al., [16] improved PLCC to 84.51% using the Padmanaban et al. video data. Our proposed approach achieved an accuracy of 87.77% using eye-tracking and head-tracking data, which is slightly less than the using EEG data and significantly better than using external physiological data [29].

The primary objective of cybersickness prediction is to know when the user is feeling discomfort during VR immersion and dynamically apply reduction techniques to alleviate cybersickness-related discomfort [33, 58]. Researchers have long been trying to predict cybersickness severity from bio-physiological signals and achieved significant accuracy in predicting the onset of cybersickness from bio-physiological signals [26, 16, 10]. However, since bio-physiological sensors data collection relies on external sensors and often get corrupted by locomotion and 3D-object manipulation, most of the prior research was only conducted in seated conditions with limited 3D-object manipulation [16, 10, 26], which is not an ideal representation of widely available VR simulations (e.g., many VR games require 3D object manipulation and free locomotion). In contrast, using an HMD's integrated sensors does not suffer from these limitations. Inspired by the above-mentioned limitations, Hewlett-Packard (HP) has recently released a commercial VR HMD (i.e., Reverb G2 Omnicept Edition), which includes HR and Eye-tracking sensors. We firmly believe that using integrated sensors is highly suitable for standalone real-time cybersickness prediction and reduction framework and other adaptive VR frameworks [33, 59].

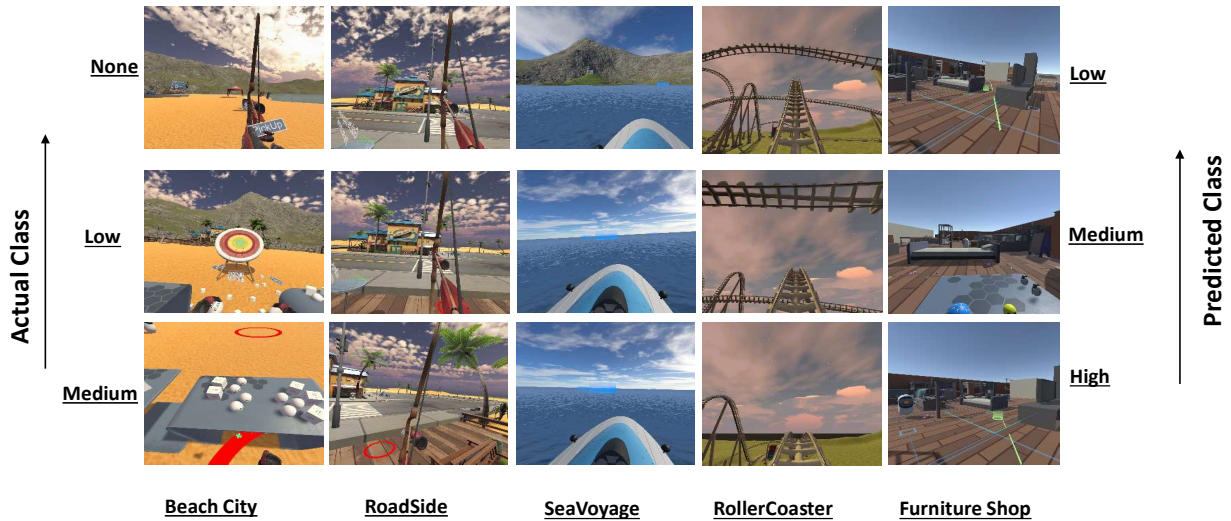

Figure 7: An misclassification example of the video data by the model. The proposed model was unable to capture the temporal features from the video data.

## 9 Limitations

Although our proposed deep fusion model achieved significant performance improvement by using eye-tracking and head-tracking data, the model performed poorly on stereoscopic video data (Table 6 and 7). The 3D-CNN model could not learn the temporal features from the stereoscopic video, optical flow, and disparity map. A misclassification example is illustrated in Figure 7. We had a total of 1755 stereoscopic videos captured from the five VR simulations. The video clips are different concerning temporal features because every participant did the task differently at a different time during each VR simulation. However, there are potential spatial similarities between the video clips. It is likely that the model only learned the spatial features of the video clips and ignoring the temporal features. In this research, we only used the SET dataset that we proposed. The model needs to be evaluated with the video datasets that are available [15]. In addition, for ground-truth construction (Figure 4), we used only $FMS_t$ reported by the participants at time $t$, which is likely to be suffered from the individual difference among the participants. Ideally, the 10-fold cross-validation should be performed with respect to the number of participants on each fold. However, this would require a large number of participants (e.g., typically more than 1000 participants) to to learn individual biases [15]. Therefore we used 10-fold cross validation similar to prior studies for cybersickenss prediction [15, 12, 16].

## 10 Conclusion and Future works

The proposed deep fusion model with eye-tracking and head-tracking data achieved an accuracy of 87.77% in cybersickness prediction. The model also performed well according to the other performance evaluation metrics (i.e., RMSE, PLCC). To the best of our knowledge, this is the state-of-the-art performance for predicting cybersickness using eye-tracking and head-tracking data. Since most of the current commercial head-mounted displays are integrated with eye and head tracking sensors, we believe that a similar approach can help develop a standalone cybersickness prediction and reduction framework [33]. We intend to explore the efficacy of using standalone cybersickness predictor from the integrated HMD sensors and test the standalone model with additional user studies in the future. We also plan to improve the ground-truth construction by utilizing SSQ scores collected after each simulation [15]. Since the model did not have expected performance with the video data, in the future, we also want to improve the performance of the stereoscopic video data by using a deep convolutional graph network [60].

## References

[1] Joshua Ratcliff, Alexey Supikov, Santiago Alfaro, and Ronald Azuma. Thinvr: Heterogeneous microlens arrays for compact, 180 degree fov vr near-eye displays. *IEEE transactions on visualization and computer graphics*, 26(5):1981–1990, 2020.

[2] Lawrence J Hettinger and Gary E Riccio. Visually induced motion sickness in virtual environments. *Presence: Teleoperators & Virtual Environments*, 1(3):306–310, 1992.

[3] Kay M. Stanney, Robert S. Kennedy, and Julie M. Drexler. Cybersickness is not simulator sickness. *Proceedings of the Human Factors and Ergonomics Society Annual Meeting*, 41(2):1138–1142, 1997.

[4] Joseph J. LaViola, Jr. A discussion of cybersickness in virtual environments. *SIGCHI Bull.*, 32(1):47–56, January 2000.

[5] Polona Caserman, Augusto Garcia-Agundez, Alvar Gámez Zerban, and Stefan Göbel. Cybersickness in current-generation virtual reality head-mounted displays: systematic review and outlook. *Virtual Reality*, pages 1–18, 2021.

[6] Simon Davis, Keith Nesbitt, and Eugene Nalivaiko. A systematic review of cybersickness. In *Proceedings of the 2014 Conference on Interactive Entertainment*, IE2014, pages 8:1–8:9, New York, NY, USA, 2014. ACM.

[7] Rifatul Islam. A deep learning based framework for detecting and reducing onset of cybersickness. In *2021 IEEE Conference on Virtual Reality and 3D User Interfaces Abstracts and Workshops (VRW)*, pages 559–560. IEEE, 2021.

[8] Susan Bruck and Paul A Watters. The factor structure of cybersickness. *Displays*, 32(4):153–158, 2011.

[9] Young Youn Kim, Hyun Ju Kim, Eun Nam Kim, Hee Dong Ko, and Hyun-Taek Kim. Characteristic changes in the physiological components of cybersickness. *Psychophysiology*, 42:616–25, 10 2005.

[10] Rifatul Islam, Yonggun Lee, Mehrad Jaloli, Imtiaz Muhammad, Dakai Zhu, Paul Rad, Yufei Huang, and John Quarles. Automatic detection and prediction of cybersickness severity using deep neural networks from user's physiological signals. In *2020 IEEE International Symposium on Mixed and Augmented Reality (ISMAR)*, pages 400–411. IEEE, 2020.

[11] Jinwoo Kim, Woojae Kim, Heeseok Oh, Seongmin Lee, and Sanghoon Lee. A deep cybersickness predictor based on brain signal analysis for virtual reality contents. In *Proceedings of the IEEE/CVF International Conference on Computer Vision*, pages 10580–10589, 2019.

[12] Hak Gu Kim, Wissam J. Baddar, Heoun-taek Lim, Hyunwook Jeong, and Yong Man Ro. Measurement of exceptional motion in vr video contents for vr sickness assessment using deep convolutional autoencoder. In *Proceedings of the 23rd ACM Symposium on Virtual Reality Software and Technology*, VRST '17, pages 36:1–36:7, New York, NY, USA, 2017. ACM.

[13] Eunhee Chang, Hyun Taek Kim, and Byounghyun Yoo. Predicting cybersickness based on user's gaze behaviors in hmd-based virtual reality. *Journal of Computational Design and Engineering*, 8(2):728–739, 2021.

[14] Benjamin Arcioni, Stephen Palmisano, Deborah Apthorp, and Juno Kim. Postural stability predicts the likelihood of cybersickness in active hmd-based virtual reality. *Displays*, 58:3–11, 2019.

[15] N. Padmanaban, T. Ruban, V. Sitzmann, A. M. Norcia, and G. Wetzstein. Towards a machine-learning approach for sickness prediction in 360° stereoscopic videos. *IEEE Transactions on Visualization and Computer Graphics*, 24(4):1594–1603, April 2018.

[16] T. M. Lee, J. Yoon, and I. Lee. Motion sickness prediction in stereoscopic videos using 3d convolutional neural networks. *IEEE Transactions on Visualization and Computer Graphics*, 25(5):1919–1927, May 2019.

[17] Heeseok Oh, Sewoong Ahn, Sanghoon Lee, and Alan Conrad Bovik. Deep visual discomfort predictor for stereoscopic 3d images. *IEEE Transactions on Image Processing*, 27(11):5420–5432, 2018.

[18] Hyunwook Jeong, Hak Gu Kim, and Yong Man Ro. Visual comfort assessment of stereoscopic images using deep visual and disparity features based on human attention. In *2017 IEEE International Conference on Image Processing (ICIP)*, pages 715–719. IEEE, 2017.

[19] Dante Risi and Stephen Palmisano. Effects of postural stability, active control, exposure duration and repeated exposures on hmd induced cybersickness. *Displays*, 60:9–17, 2019.

[20] Miguel Melo, José Vasconcelos-Raposo, and Maximino Bessa. Presence and cybersickness in immersive content: Effects of content type, exposure time and gender. *Computers & Graphics*, 71:159–165, 2018.

[21] Behrang Keshavarz and Heiko Hecht. Validating an efficient method to quantify motion sickness. *Human Factors*, 53(4):415–426, August 2011.

[22] Robert S. Kennedy, Norman E. Lane, Kevin S. Berbaum, and Michael G. Lilienthal. Simulator sickness questionnaire: An enhanced method for quantifying simulator sickness. *The International Journal of Aviation Psychology*, 3(3):203–220, 1993.

[23] S. V. G. Cobb, S. Nichols, A. Ramsey, and J. R. Wilson. Virtual reality-induced symptoms and effects (vrise). *Presence*, 8(2):169–186, 1999.

[24] Eugene Nalivaiko, Simon L. Davis, Karen L. Blackmore, Andrew Vakulin, and Keith V. Nesbitt. Cybersickness provoked by head-mounted display affects cutaneous vascular tone, heart rate and reaction time. *Physiology & Behavior*, 151:583 – 590, 2015.

[25] Hiroshi Oyamada, Atsuhiko Iijima, Akira Tanaka, Kazuhiko Ukai, Haruo Toda, Norihiro Sugita, Makoto Yoshizawa, and Takehiko Bando. A pilot study on pupillary and cardiovascular changes induced by stereoscopic video movies. *Journal of neuroengineering and rehabilitation*, 4:37, 02 2007.

[26] Yi-Tien Lin, Yu-Yi Chien, Hsiao-Han Wang, Fang-Cheng Lin, and Yi-Pai Huang. 65-3: The quantization of cybersickness level using eeg and ecg for virtual reality head-mounted display. *SID Symposium Digest of Technical Papers*, 49(1):862–865, 2018.

[27] Alireza Mazloumi Gavgani, Keith V. Nesbitt, Karen L. Blackmore, and Eugene Nalivaiko. Profiling subjective symptoms and autonomic changes associated with cybersickness. *Autonomic Neuroscience*, 203:41 – 50, 2017.

[28] Yoichi Yokota, Mitsuhiro Aoki, Keisuke Mizuta, Yatsuji Ito, and Naoki Isu. Motion sickness susceptibility associated with visually induced postural instability and cardiac autonomic responses in healthy subjects. *Acta oto-laryngologica*, 125(3):280–285, 2005.

[29] Mark S. Dennison, A. Zachary Wisti, and Michael D'Zmura. Use of physiological signals to predict cybersickness. *Displays*, 44:42 – 52, 2016. Contains Special Issue Articles – Proceedings of the 4th Symposium on Liquid Crystal Photonics (SLCP 2015).

[30] D. Jeong, S. Yoo, and J. Yun. Cybersickness analysis with eeg using deep learning algorithms. In *2019 IEEE Conference on Virtual Reality and 3D User Interfaces (VR)*, pages 827–835, March 2019.

[31] Marije Jansen, Thomas P White, Karen J Mullinger, Elizabeth B Liddle, Penny A Gowland, Susan T Francis, Richard Bowtell, and Peter F Liddle. Motion-related artefacts in eeg predict neuronally plausible patterns of activation in fmri data. *Neuroimage*, 59(1):261–270, 2012.

[32] Shahab Ardalan, Siavash Moghadami, and Samira Jaafari. Motion noise cancelation in heartbeat sensing using accelerometer and adaptive filter. *IEEE Embedded Systems Letters*, 7(4):101–104, 2015.

[33] Rifatul Islam, Samuel Ang, and John Quarles. Cybersense: A closed-loop framework to detect cybersickness severity and adaptively apply reduction techniques. In *2021 IEEE Conference on Virtual Reality and 3D User Interfaces Abstracts and Workshops (VRW)*, pages 148–155. IEEE, 2021.

VR Sickness Prediction from Integrated HMD's Sensors using Multimodal Deep Fusion Network

[34] Andrej Karpathy, George Toderici, Sanketh Shetty, Thomas Leung, Rahul Sukthankar, and Li Fei-Fei. Large-scale video classification with convolutional neural networks. In *Proceedings of the IEEE conference on Computer Vision and Pattern Recognition*, pages 1725–1732, 2014.

[35] Jonathan Masci, Ueli Meier, Dan Cireşan, and Jürgen Schmidhuber. Stacked convolutional auto-encoders for hierarchical feature extraction. In Timo Honkela, Włodzisław Duch, Mark Girolami, and Samuel Kaski, editors, *Artificial Neural Networks and Machine Learning – ICANN 2011*, pages 52–59, Berlin, Heidelberg, 2011. Springer Berlin Heidelberg.

[36] Andreas Geiger, Philip Lenz, and Raquel Urtasun. Are we ready for autonomous driving? the kitti vision benchmark suite. In *2012 IEEE Conference on Computer Vision and Pattern Recognition*, pages 3354–3361. IEEE, 2012.

[37] Weina Jin, Jianyu Fan, Diane Gromala, and Philippe Pasquier. Automatic prediction of cybersickness for virtual reality games. 08 2018.

[38] Phil Lopes, Nana Tian, and Ronan Boulic. Eye thought you were sick! exploring eye behaviors for cybersickness detection in vr. In *Motion, Interaction and Games*, pages 1–10. 2020.

[39] Cuiting Guo, Jennifer Ji, and Richard So. Could okan be an objective indicator of the susceptibility to visually induced motion sickness? In *2011 IEEE Virtual Reality Conference*, pages 87–90. IEEE, 2011.

[40] Tobias Feigl, Daniel Roth, Stefan Gradl, Markus Wirth, Marc Erich Latoschik, Bjoern M Eskofier, Michael Philippsen, and Christopher Mutschler. Sick moves! motion parameters as indicators of simulator sickness. *IEEE transactions on visualization and computer graphics*, 25(11):3146–3157, 2019.

[41] Christopher Widdowson, Israel Becerra, Cameron Merrill, Ranxiao Frances Wang, and Steven LaValle. Assessing postural instability and cybersickness through linear and angular displacement. *Human factors*, page 0018720819881254, 2019.

[42] Stéphane Bouchard, Geneviève Robillard, and Patrice Renaud. Revising the factor structure of the simulator sickness questionnaire. *Annual Review of CyberTherapy and Telemedicine*, 5:128–137, 01 2007.

[43] Jann Philipp Freiwald, Yvonne Göbel, Fariba Mostajeran, and Frank Steinicke. The cybersickness susceptibility questionnaire: predicting virtual reality tolerance. In *Proceedings of the Conference on Mensch und Computer*, pages 115–118, 2020.

[44] Hyun K. Kim, Jaehyun Park, Yeongcheol Choi, and Mungyeong Choe. Virtual reality sickness questionnaire (vrsq): Motion sickness measurement index in a virtual reality environment. *Applied Ergonomics*, 69:66 – 73, 2018.

[45] Gunnar Farnebäck. Two-frame motion estimation based on polynomial expansion. In *Scandinavian conference on Image analysis*, pages 363–370. Springer, 2003.

[46] Heiko Hirschmuller. Stereo processing by semiglobal matching and mutual information. *IEEE Transactions on pattern analysis and machine intelligence*, 30(2):328–341, 2007.

[47] Keith Nesbitt, Simon Davis, Karen Blackmore, and Eugene Nalivaiko. Correlating reaction time and nausea measures with traditional measures of cybersickness. *Displays*, 48:1–8, 2017.

[48] Cees GM Snoek, Marcel Worring, and Arnold WM Smeulders. Early versus late fusion in semantic video analysis. In *Proceedings of the 13th annual ACM international conference on Multimedia*, pages 399–402, 2005.

[49] Abien Fred Agarap. Deep learning using rectified linear units (relu). *arXiv preprint arXiv:1803.08375*, 2018.

[50] Subhransu Maji, Lubomir Bourdev, and Jitendra Malik. Action recognition from a distributed representation of pose and appearance. In *CVPR 2011*, pages 3177–3184. IEEE, 2011.

[51] Yoshua Bengio and Yves Grandvalet. No unbiased estimator of the variance of k-fold cross-validation. *Journal of machine learning research*, 5(Sep):1089–1105, 2004.

[52] Juan D Rodriguez, Aritz Perez, and Jose A Lozano. Sensitivity analysis of k-fold cross validation in prediction error estimation. *IEEE transactions on pattern analysis and machine intelligence*, 32(3):569–575, 2009.

[53] Gareth James, Daniela Witten, Trevor Hastie, and Robert Tibshirani. *An introduction to statistical learning*, volume 112. Springer, 2013.

[54] Brian D Ripley. *Pattern recognition and neural networks*. Cambridge university press, 2007.

[55] Rich Caruana, Steve Lawrence, and Lee Giles. Overfitting in neural nets: Backpropagation, conjugate gradient, and early stopping. *Advances in neural information processing systems*, pages 402–408, 2001.

[56] Hak Gu Kim, Heoun-Taek Lim, Sangmin Lee, and Yong Man Ro. Vrsa net: Vr sickness assessment considering exceptional motion for 360 vr video. *IEEE transactions on image processing*, 28(4):1646–1660, 2018.

[57] Sunu Wibirama, Paulus Insap Santosa, Putu Widyarani, Nanda Brilianto, and Wina Hafidh. Physical discomfort and eye movements during arbitrary and optical flow-like motions in stereo 3d contents. *Virtual Reality*, 24(1):39–51, 2020.

[58] Y. Y. Kim, E. N. Kim, M. J. Park, K. S. Park, H. D. Ko, and H. T. Kim. The application of biosignal feedback for reducing cybersickness from exposure to a virtual environment. *Presence*, 17(1):1–16, 2008.

[59] Mohammadhossein Moghim, Robert Stone, Pia Rotshtein, and Neil Cooke. Adaptive virtual environments: A physiological feedback hci system concept. In *2015 7th Computer Science and Electronic Engineering Conference (CEEC)*, pages 123–128. IEEE, 2015.

[60] Feng Mao, Xiang Wu, Hui Xue, and Rong Zhang. Hierarchical video frame sequence representation with deep convolutional graph network. In *Proceedings of the European Conference on Computer Vision (ECCV) Workshops*, pages 0–0, 2018.